\documentclass[a4paper]{JACoW-GSI}
\usepackage{graphicx}
\usepackage{url}

\begin{document}
\title{The $K\pi$ and $\pi\pi$ S-wave from $D$ decays}

\author{A. C. dos Reis}
\affil{CBPF, Rio de Janeiro, Brazil}

\maketitle

\section{Introduction}

The physics of the scalar mesons has been challenging for decades. Scalar 
mesons are difficult to detect, since they decay isotropically and have, 
in general, large widths. There are many candidates with mass bellow 2 GeV/$c^2$. 
Some states are now well established, while others remain controversial.
In any case, there is a large overlap between states in this region of the spectrum.

An additional problem arises from the richness of the low energy strong 
dynamics, allowing other $0^{++}$ configurations than the usual $q \bar q$. 
Although none of these 'exotic' configurations has been clearly established, 
some would populate the 1-2  GeV/$c^2$ region, mixing with the regular $q \bar q$ mesons.
The identification of the $q \bar q$ scalar nonet(s) is, thus, a rather
complicated task, which can only be accomplished if one combines data from 
different types.

This paper is focused on two issues: the $K\pi$ spectrum near 
threshold -- the kappa problem --, and the $\pi\pi$ spectrum between 1.2 and 
1.5 GeV/$c^2$ -- the $f_0(1370)$ problem. Both problems will be analysed from 
the perspective of the $D$ meson decays, with some additional information from
$\tau$ and $B$ decays.

Hadronic and semileptonic decays of $D$ mesons have unique features that 
make them a key to light meson spectroscopy, in particular to the study of
the  $K\pi$ and $\pi\pi$ scattering amplitudes in S-wave.
First, these are the only process that allows us to access
the $K\pi$/$\pi\pi$ spectrum continuously from threshold up to $\sim$ 1.5 GeV/$c^2$. 
In $K\pi$ and $\pi\pi$ scattering, the production of scalar resonances near 
threshold is suppressed by the Adler zeroes. No such effect is observed in 
$D$ decays. In scattering, the large nonresonant component forms a continuum
background on top of which the scalar resonances are found. The interference
between the broad states and this continuum distorts the resonance line shape, and
is always a difficult problem
to be accounted for. In $D$ decays, the nonresonant component is usually small.
In $D$ decays, channels with two identical pions in the final state have
a largely dominant S-wave component. Decay modes like $D^+\to K^-\pi^+\pi^+$ and 
$D^+,D^+_s \to \pi^-\pi^+\pi^+$ are easy to be reconstructed and have large 
branching  fractions. There are plenty of good data with very low background
on these 'golden' modes. In the first year of the LHCb operation, we will enter
in the regime of 'infinite' statistics.

The decay of a $D$ meson is, obviously, a very complex process. It is initiated by 
the $c \to s(d)$  weak transition. This transition is embedded in a
strongly interacting system, in the non-perturbative regime, from which the final 
state hadrons emerge. No precise quantitative description based on first principles 
can be performed. 

\begin{figure}[htb]
\centering
\includegraphics*[width=65mm]{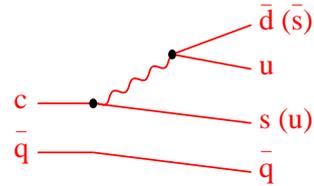}
\vskip -2cm
\caption{The $W$-radiation diagram. In the case of the $D^+ \to K^-\pi^+\pi^+$
decay ($\bar q = \bar d$) the intermediate states are dominated by
the $\overline{K}^*$ family ($s\bar d$).}
\label{diag}
\end{figure}

Nevertheless, a qualitative description of such a complex process can be constructed
using simple ideas. Going through the PDG listings, one realises that
essentially the whole $D$ decay width can be explained by
simple tree-level valence quark diagrams, such as the one shown in 
Fig. \ref{diag}, connected to the well known $q \bar q$ resonances from the 
Constituent Quark Model. Considering, for instance, three-body decays proceeding through
intermediate states having spin-1 and 2 resonances, 
one concludes that the regular $q \bar q$ mesons correspond to the entire
decay rate. No 'exotic' states have been observed in $D$ decays, which act as a 
$q \bar q$ filter. Assuming the tree-level diagrams to be dominant, the 
available 'final state' quarks, that is, those resulting from the weak $c$ decay,
determine not only which $q \bar q$ resonances can occur, but also their relative rates. 
A nice example is the 'weak vector/axial-vector dominance',  a close analogy 
to the vector dominance in electrodynamics. Due to the V-A nature of the 
$c$ quark decay, in the diagram of Fig. \ref{diag} the virtual $W^+$ will 
couple preferentially to a vector or to an axial-veto particle, rather than 
to a pseudoscalar. The branching fraction for the decays 
$D^0 \to K^-a_1^+(1260)$ and $D^0 \to K^-\rho^+(770)$, for instance, 
are three to four times larger than that of $D^0 \to K^-\pi^+$.

\section{On the analysis technique}

Essentially all studies of three-body hadronic $D$ decays employ the
same technique: the unbinned maximum likelihood fit of the Dalitz plot, 
in which the decay matrix element is represented by a coherent sum of
phenomenological amplitudes \cite{pdg}. These amplitudes correspond to the possible 
intermediate states in the decay chain $D \to Rh$, $R\to hh$ ($h=K,\pi$). The amplitudes
are grouped according to the orbital angular momentum $L$ in the first step of
the decay chain, 

\[
S_{\mathrm{pdf}} = \mid \sum_L A_L \mid^2, \ A_L = \sum_k c^L_k \mathcal{A}^L_k  
\]

The amplitudes  \ $\mathcal{A}^L_k$  \ are weighted by constant complex 
coefficients $c_k^L$, the series being truncated at $L=2$. 
The set of complex coefficients is, in general, the fit output. 

In the case of a resonance with spin, the 
standard procedure is to define the resonant amplitude $\mathcal{A}_k$ as 
a product of a relativistic Breit-Wigner function, form factors (usually the 
Blatt-Weisskopf dumping factors \cite{blatt}) for the $D$ and the 
resonance decay vertexes and a function describing the angular distribution of 
the final state particles, accounting for the angular momentum conservation.
The  S-wave is the problematic issue. The 
way it is handled consists in the basic difference between the various 
Dalitz plot analyses. Here I will briefly describe the most common 
approaches to the S-wave.

\subsection{The Isobar Model}

In the so called Isobar Model the S-wave is usually assumed to be a sum
of a constant  nonresonant term and Breit-Wigner functions for the scalar resonances. 
The Breit-Wigner functions may or may not be multiplied by scalar form 
factors. In spite of conceptual problems, in most cases the  Isobar Model
provides a reasonably good, effective description of the data. 

In the beginning of this decade the pioneer work of the E791 Collaboration \cite{ds,dp,cg}
showed evidence for two broad scalar resonances, identified to 
the $\sigma$ and the to $\kappa$ in the study of the $D^+ \to \pi^-\pi^+\pi^+$ and 
$D^+ \to K^-\pi^+\pi^+$ decays, respectively. The E791 analysis used
the Isobar Model, with one innovation: in addition to the constant complex 
coefficients $c_k^L$, the masses and widths of the resonances were also determined by
the fit.

The $\sigma$ and the $\kappa$, nowadays
well established states, were soon after confirmed by other experiments, in
different reactions and with higher statistics \cite{cleosig,kmat1,bessig,beskap}.
The values obtained by E791
for the Breit-Wigner masses and widths are inadequate for determining 
the $\sigma$ and $\kappa$ poles. The merit of the E791 work was to demonstrate 
the existence of structures at low $\pi^+\pi^+$ and $K^-\pi^+$ mass 
with a resonant behaviour, that is, described only by an amplitude with a complex, 
energy-dependent phase. The Breit-Wigner was the simpler form of such an
amplitude. 

In spite of yielding a good description of the data,
the Isobar Model has a limited ability in disentangling individual contributions
from broad components in the S-wave. The case of the $D^+ \to K^-\pi^+\pi^+$ 
decay is typical: the $\kappa$ and the nonresonant components are so highly correlated
that the determination of the decay fractions become rather uncertain.

\begin{figure}[htb]
\centering
\includegraphics*[width=65mm]{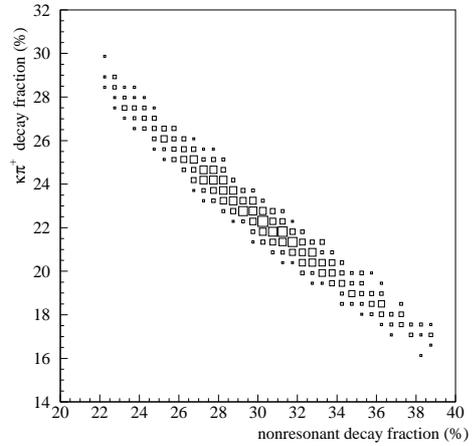}
\caption{The correlation between the $\kappa \pi^+$ and the non-resonant
amplitudes from the Dalitz plot analysis of the $D^+ \to K^-\pi^+\pi^+$ using 
the isobar model. See text for details.}
\label{corr}
\end{figure}

This readily illustrated by the following exercise based on the result of the
Isobar fit of $D^+ \to K^-\pi^+\pi^+$ Dalitz plot from FOCUS
\cite{kmat}. The set of coefficients from the FOCUS fit (Table II of
ref. \cite{kmat} is taken as the input model to simulate an
ensemble with 2000 Dalitz plots. Each simulated Dalitz plot 
had the same number (54K) of signal events as in the FOCUS data set.
If there were no statistical fluctuations of the signal distribution,
these 2000 samples would be identical.
Each Dalitz plot was fitted with the same model. The resulting decay fractions were recorded. 
A scatter plot of  the $\kappa \pi^+$ and nonresonant decay fractions from
the 2000 fits is displayed in Fig. \ref{corr}. In the absence of
correlations, the projection of the scatter plot onto each axis should look like a
Gaussian centred at the value quoted in Table II of \cite{kmat}; the width
should match the statistical error from the FOCUS data fit. 
But what we see is that the correlation between these two amplitudes is 
indeed very high, showing that the Isobar Model cannot provide a reliable 
distinction between the broad structures of the S-wave.

\subsection{The K-matrix approach}

\begin{figure}[htb]
\centering
\includegraphics*[width=65mm]{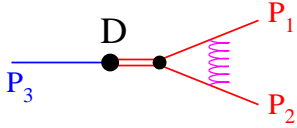}
\vskip -2cm
\caption{Schematic diagram of a three-body decay of a $D$ meson in which the resonant system
$P_1P_2$ does not interact with the third particle. This is the underlying picture
in the K-matrix approach. According to it final state interaction occurs only between 
$P_1$ and $P_2$.}
\label{fsi2}
\end{figure}

An alternative approach to the S-wave is the K-matrix formalism,
applied to Dalitz plot analyses of $D$ decays by the FOCUS
collaboration \cite{kmat,kmat1}. This approach involves a very sophisticated machinery,
but is based on a unrealistic and somewhat naive assumption: in the three-body final state,
the $\pi^+\pi^-$/$K^-\pi^+$ pair forms an isolated system, which evolves as if the 
third body was not there. Three-body interaction is, therefore, ignored, as shown 
schematically in Fig. \ref{fsi2}. 

Given that a no rigorous treatment of a three-body final
state strong interaction exists, this assumption greatly simplifies
the problem. Only under this assumption one can invoke arguments such as
two-body unitarity. In the absence of a full three-body final state interaction (FSI),
the dynamics of the $\pi^-\pi^+\pi^+$ and $K^-\pi^+\pi^+$ final states 
becomes entirely determined by the two-body  $\pi^+\pi^-$ and $K^-\pi^+$ 
interactions, respectively. That allows one to constrain the $D$ decay amplitude by data
from different reactions. The phase of the $\pi^+\pi^-$/$K^-\pi^+$ 
amplitude should, therefore, match that of the $\pi^+\pi^-$/$K^-\pi^+$ 
scattering not only for the S-wave, but also for all other waves. 
That is the essence of Watson's theorem.

There is no experimental evidence supporting this approximation. As we will
see, the S- and P-wave phases from $D$ decays are rather different from that of
$\pi^+\pi^-$/$K^-\pi^+$ scattering \cite{brian,eu,babar3pi}.

\vskip -2cm

In the K-matrix approach the  $\pi^+\pi^-$/$K^-\pi^+$ S-wave phase
is fixed. The K-matrix approach is, hence, not intended to provide
new data on the $\pi^+\pi^-$/$K^-\pi^+$ S-wave phase. The decay amplitude is
defined as a product of the K-matrix and a production amplitude. The  fit parameters are
contained in the production amplitude. That includes an adjustable energy dependent phase. 
In other words, the observed phase from $D$ decays is modeled by the sum of the
known phase from scattering and an unknown phase to be determined by the fit.
The 'production phase' can, therefore, account for any eventual differences between
the S-wave phase form $\pi^+\pi^-$/$K^-\pi^+$ scattering and from $D$ decays. With
such a freedom, the 
K-matrix approach provides, in general, fits with acceptable quality.

\subsection{The MIPWA method}

The Model Independent Partial Wave Analysis technique was developed by the
E791 Collaboration \cite{brian}, implementing an idea put forward by W.
Dunwoodie. As in the isobar and K-matrix approaches, the decay matrix element 
is written as a sum of partial waves, truncated at the D-wave (which is already a 
very small contribution).
No assumption is made on the nature of the S-wave, which is
represented by a generic complex function to be determined directly from data,

\begin{equation}
A_0(s) = a_0(s)e^{i\phi_0(s)}.
\label{c0}
\end{equation}

The $\pi^+\pi^-$/$K^-\pi^+$ mass spectrum is divided into $n$ slices ($n>20$, 
in general). For each slice two real numbers are fitted,
so that at the {\em k-th} slice $A_0(s=s_k)=a_0^k e^{i\phi_0^k}$. 
An interpolation is used to define the 
value of the S-wave in any point between $s_k\leq s < s_{k+1}$. The set of $\{a_0^k,\phi_0^k\}$, together
with the coefficients $c_k^L$ are the fit parameters.

\begin{figure}[htb]
\centering
\includegraphics*[width=65mm]{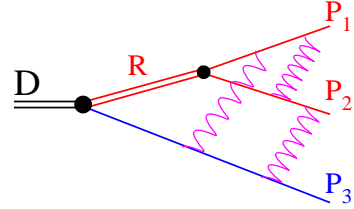}
\vskip -2cm
\caption{A more realistic schematic diagram of a $D$ decay. Final state interactions
may occur between all decay particles, including three-body interactions.}
\label{fsi3}
\end{figure}

In this sense, the MIPWA method is the most exempt approach. The only
assumption is common to all other analyses, that is, the P- and D-waves 
are well described by a sum of Breit-Wigner amplitudes. The are some shortcuts,
though. First, one should handle a large number of fit parameters (the S-wave alone
has 2$n$ free parameters), which introduces some technical difficulties.
Moreover, the MIPWA S-wave relies on a precise representation 
of the other waves. If something is wrong with the P- and D-wave parametrisation,  
their content would "leak" into the S-wave. But the crucial problem is that the MIPWA 
S-wave is an inclusive measurement, since the $\pi^+\pi^-$/$K^-\pi^+$ system is
embedded in a three-body strongly interacting final state, as illustrated in
Fig \ref{fsi3}. Extracting the pure $\pi^+\pi^-$/$K^-\pi^+$ amplitude is not a
trivial task: one needs to deconvolute the desired phase from the ones introduced by
three-body FSI \cite{pennington} and, perhaps, from the production amplitude.

\vskip -1cm


\section{The $K\pi$ amplitude -- the $\kappa$ problem}

The existence of a broad scalar resonance at low $K^-\pi^+$ mass was first reported 
by the E791 Collaboration \cite{cg} from the Dalitz plot analysis of the
$D^+ \to K^-\pi^+\pi^+$. This state was identified with the $\kappa(800)$ meson. Shortly after
the same state was observed by several other experiments \cite{beskap,kmat,CLEO-c}.

The the nature of the $\kappa(800)$ meson -- an $I=1/2$ state? -- has been the subject of a 
long-standing debate.  While there is now plenty of evidence for the neutral state,
results for the charged partner are still scarce and conflicting \cite{babarkkpi0,belletau}. 

The $\kappa(800)$ pole position has been determined recently using LASS data \cite{lass} and a
Roy-Steiner representation of  $K^-\pi^+$ scattering amplitude \cite{dg}. Note, however, that 
there is no data on $K^- \pi^+ \to K^- \pi^+$ 
bellow 825 MeV/$c^2$, where LASS data starts. The crucial issues are, therefore: a) to fill the 
existing gap between the $K^- \pi^+$ threshold and 825 MeV/$c^2$; b) to find the charged
$\kappa(800)$ state.

Searching for the charged $\kappa$ is a much harder task than for the neutral partner. There is
no 'golden mode' in which the contribution of the $(K\pi)^{\pm}$ in S-wave is largely dominant.
The available data sets have still limited statistics. This will be illustrated by two studies,
with a somewhat surprising results.

\subsection{$D^0 \to K^-K^+\pi^0$ \hskip .2cm from BaBar}

The $D^0 \to K^-K^+\pi^0$ decay was studied by BaBar \cite{babarkkpi0}. The Babar sample has
11000 events with 98\% purity. The $D^0 \to K^-K^+\pi^0$ is a Cabibbo suppressed 
decay, with dominant tree-level amplitudes (external and internal {\em W}-radiation). The
dominant contributions should come from the vector modes $\bar K^{*+}K^-$ and $\phi \pi^0$, but
we also expect sizable decay fractions for the modes  $\bar K^{*-}K^+$ and 
$(K\pi)_S^{\pm}K^{\mp}$. The $K\pi$ S-wave can be analysed from threshold up to 1.4 GeV/$c^2$ 
using the $K^{\pm}\pi^0$ system. 

\begin{figure}[htb]
\centering
\includegraphics*[width=70mm]{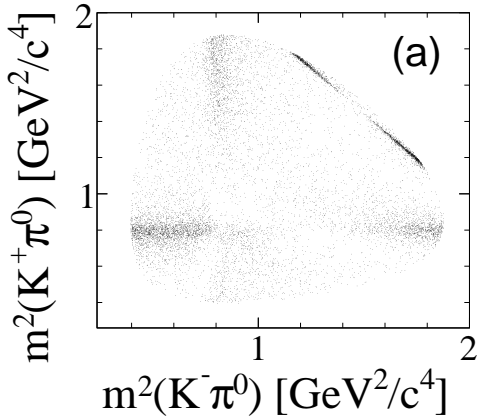}
\caption{The Dalitz plot of the $D^0 \to K^-K^+\pi^0$ decay. The narrow band on the top right
part of the plot corresponds to the $\phi\pi^0$ mode. The horizontal structure is due
to the dominant mode, the  $K^{*+}K^-$. The mode $\overline{K}^{*-}K^+$ appears as a vertical
band. In all cases we see the node due to the angular distribution typical of a spin-1 resonance.}
\label{kkpi0}
\end{figure}

\begin{figure}[htb]
\centering
\includegraphics*[width=40mm]{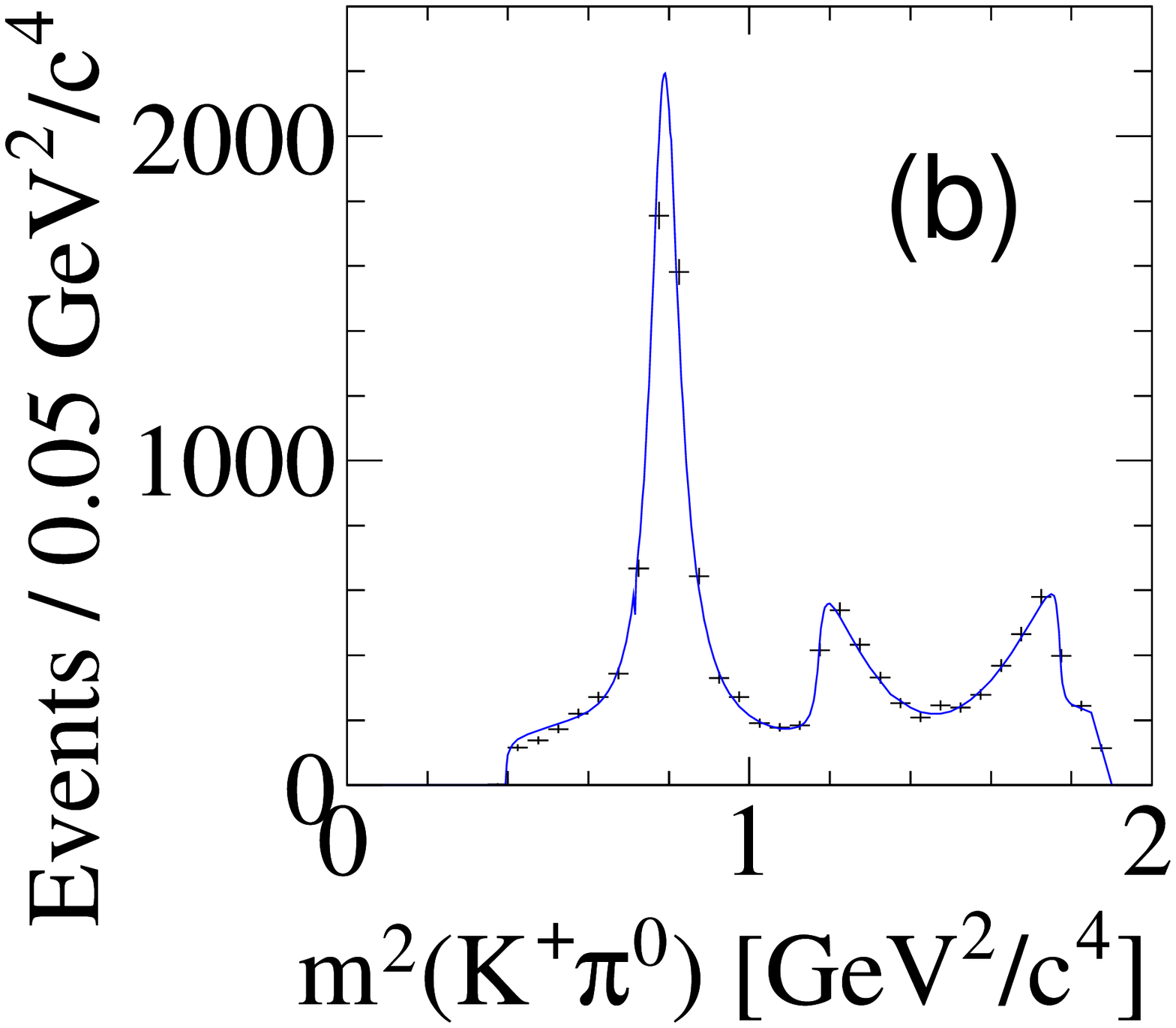} \includegraphics*[width=40mm]{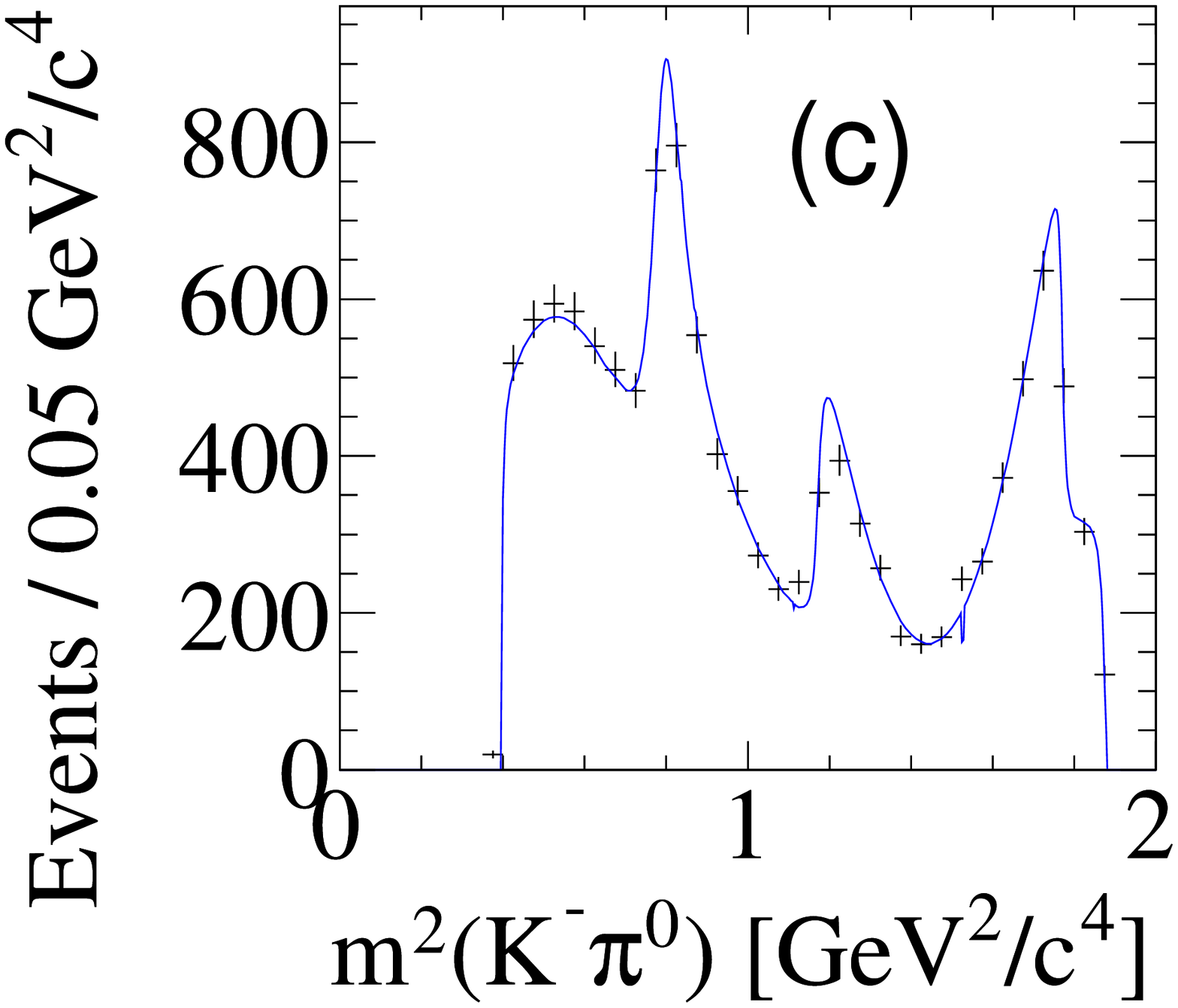}
\caption{Projection of the $D^0 \to K^-K^+\pi^0$ Dalitz plot of Fig. \ref{kkpi0}. The projection
on the $K^+\pi^0$ axis (left plot) show the peak of the $K^{*}(892)^{+}$, much more prominent than
the one in the $\overline{K}^-\pi^0$ projection (right plot). In both plots the structure between 
1-2 GeV$^2$/$c^4$ is the projection of the $\phi\pi^0$ mode.}
\label{kkpi0b}
\end{figure}

The Dalitz plot of the $D^0 \to K^-K^+\pi^0$ decay is shown in Fig. \ref{kkpi0}. We see clearly
the bands corresponding to the $\bar K^{*+}K^-$,  $\bar K^{*-}K^+$ and $\phi \pi^0$ modes
(the $KK$ axis runs along the top-right to bottom-left diagonal). The
Dalitz plot projections are shown in Fig. \ref{kkpi0b}, confirming the expectation of a
larger $\bar K^{*+}K^-$ contribution compared to $\bar K^{*-}K^+$.

The Dalitz plot of Fig. 5 was fit with three different models for the $K^{\pm}\pi^0$ S-wave. In the first 
model, the S-wave was represented by the LASS $I$=1/2 amplitude,

\begin{equation}
A_0(s) = \frac{\sqrt{s}}{p}\sin \delta(s) e^{i \delta(s)} \ ,
\label{effrange1}
\end{equation}
where

\begin{equation}
\delta(s) = \cot^{-1} \left ( \frac{1}{pa} + \frac{bp}{2} \right ) + 
\cot^{-1} \left ( \frac{M_0^2-s}{M_0\Gamma_0 \frac{M_0}{\sqrt{s}}\frac{p}{p_0}} \right )
\label{effrange2}
\end{equation}

In the above equation $s$ is the $K\pi$ mass squared, $a$ and $b$ are real constants
and $p$ is the kaon momentum in the
$K\pi$ rest frame. The parameters  $M_0$ and $\Gamma_0$ refer to the $K^{*}_0(1430)$ resonance.

In the second model BaBar used the E791 MIPWA $K^-\pi^+$ amplitude \cite{brian}. In
the third model a coherent sum of a uniform nonresonant term and Breit-Wigner amplitudes for
the $\overline{K}^{*\pm}_0(1430)K^{\mp}$ and $\kappa(800)^{\pm}K^{\mp}$ modes -- the isobar model. 

The isobar model yielded the smaller fit probability. The best fit was obtained with the LASS
$I$=1/2 S-wave amplitude, although a good fit was also achieved with E791 MIPWA S-wave. The latter
describes the data well, except in the region near threshold.

BaBar data is well described by two models, both using the LASS $I$=1/2 
amplitude for the $K^{\pm}\pi^0$ S-wave.  The first model has nine amplitudes, 
whereas the second has only six. The basic difference between the
two models are the  $K^{*}(1410)^{\pm}K^{\mp}$ modes, present in the first model and absent in
the second one. The decay fractions from the two fits are listed in Table \ref{kpipi0}.

The failure of the isobar model in describing the data cannot
be taken as an argument against the $\kappa$. Recall that the $\kappa$ pole was found in
LASS data. One would expect the  LASS amplitude to be well suited for
situations where the $K\pi$ system is isolated from the rest of the final state,
as in semileptonic decays. The fact that the LASS amplitude yielded the best fit is
a bit surprising.  

We should analyse this result with some care, though.
An inspection of the BaBar fit fractions in Table I shows that the
interpretation is not straightforward. The decay fractions from model II, where the
tiny contributions of the $K^{*}(1410)^{\pm}K^{\mp}$ modes were removed,
add up to 165\%! The fraction of the $K^{*}(1410)$ resonance in $D$ decays
is always marginal, as in the present case. However, when this small component is removed
from the fit the P-wave remains unaltered, but the
$K^{+}\pi^0$ S-wave contribution jumps from 16.3\% to
71.1\%. The $K^-\pi^0$ S-wave component, consistent with zero in model I, becomes a 10$\sigma$ effect
in model II. There is an obvious interplay between the $K^*(1410)K$ and the $K^{\pm}\pi^0$ S-wave.
With more data the $K^{+}\pi^0$ S-wave amplitude could be extracted with the MIPWA
technique.

\begin{table}
\begin{tabular}{ccc}   \hline 
 mode                &    model I   &  model II    \\ \hline

 $K^*(892)^+K^-$     & $45.2\pm0.9$ & $44.4\pm0.9$ \\  

 $K^*(1410)^+K^-$    & $3.7\pm1.5$  &  -           \\

 $K^+\pi^0 (S) $     & $16.3\pm0.1$ & $71.1\pm4.2$ \\ 

 $\phi \pi^0$        & $19.3\pm0.7$ & $19.4\pm0.7$ \\

 $f_0(980)\pi^0$     & $6.7\pm1.8$  & $10.5\pm1.4$ \\

 $K^*(892)^-K^+$     & $16.0\pm0.9$ & $15.9\pm0.9$ \\  

 $K^*(1410)^-K^+$    & $2.7\pm1.5$  &  -           \\ 
 
 $K^-\pi^0 (S) $     & $2.7\pm1.5$ & $3.9\pm1$ \\  \hline
\end{tabular}
\caption{Decay fractions, in \%, from the BaBar $D^0 \to K^-K^+\pi^0$ Dalitz plot fit using 
the LASS  $I$=1/2 amplitude for the $K^{\pm}\pi^0$ S-wave . As expected, the 
$K^*(892)^+K^-$ is the dominant mode. Removing the small $K^*(1410)K$ component causes
a drastic change in the $K^+\pi^0$ S-wave contribution, bringing the sum of the decay
fractions to over 165\%.}
\label{kpipi0}
\end{table}

\subsection{$\tau^- \to \overline{K}^0\pi^-\nu_{\tau}$ \hskip .2cm  from Belle}

Semileptonic decays like $D\to K\pi l\nu$ and $\tau \to K\pi \nu$  are very 
interesting alternatives, since the $K\pi$ system is free from final state strong interaction.
We should expect Watson's theorem to hold, or, in other words, that the $K\pi$ S-wave phase 
matches that from LASS. There are some problems with semileptonic decays, though. In these
decays the P-wave corresponds to over 90\% of the decay rate. Very large samples are required
in order to have a reasonable statistics for the S-wave.

The $\tau^- \to \overline{K}^0\pi^-\nu_{\tau}$ decay was studied by Belle \cite{belletau}.
The sample was selected from events of the type \ $e^+e^- \to \tau^+\tau^-$, 
with \ $\tau^+ \to l^+ \nu_{\tau} \nu_l$ \
and  \ $\tau^- \to K_S\pi^- \nu_{\tau}$. The signature was a lepton recoiling against a pair of pions
of opposite charge. The selected sample has 53K signal events.

\begin{figure}[htb]
\centering
\includegraphics*[width=70mm]{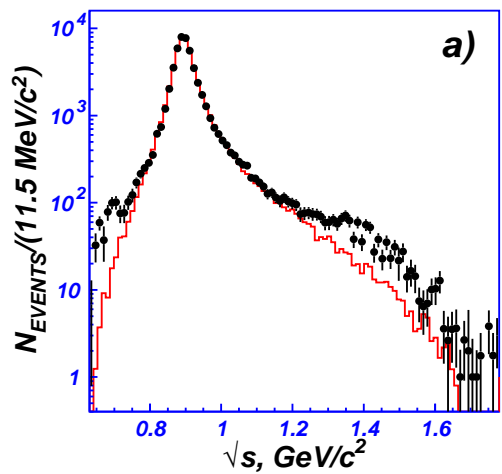}
\caption{The $K_s\pi^-$ spectrum from the $\tau^- \to \overline{K}^0\pi^-\nu_{\tau}$ decay, 
from Belle. The solid histogram is the contribution of the $\overline{K}^{*}(892)^-$.}
\label{belletau1}
\end{figure}

The $K_s\pi^-$ mass spectrum is shown in Fig. \ref{belletau1}. Superimposed (histogram in red)
we see the $\overline{K}^{*}(892)^-$ contribution. There is an excess of data events over the
$\overline{K}^{*}(892)^-$ contribution both at 
the lower and the higher part of the $K_s\pi^-$ spectrum. The spectrum of Fig. \ref{belletau1}
was fitted with different models. To the
dominant $\overline{K}^{*}(892)^-$ two other amplitudes were added: the $\kappa(800)^-$ plus one 
$K\pi$ resonance with higher mass --- either the $\overline{K}^*(1410)^-$, $\overline{K}^*_0(1430)^-$ or
the $\overline{K}^*(1680)^-$. The LASS amplitude (eq. \ref{effrange1} and \ref{effrange2}) was also
tried.

The result of the fit of the $K_s\pi^-$ spectrum was also surprising. 
Contrarily to what one would expect, the model with the LASS $I$=1/2 amplitude fails
to reproduce the $K_s\pi^-$ line shape (C.L.=$10^{-8}$). The best description of the
data was achieved by adding to the $\overline{K}^{*}(892)^-$ a pure scalar component, that is, 
the model with the $\kappa(800)^-$ plus the $\overline{K}^*_0(1430)^-$ resonance. 

The missing neutrinos introduce additional limitations. The full event reconstruction becomes
very difficult. One has to handle a relatively high background, at the 20\% level
in this analysis. The most serious consequence is that, since the position of both the primary and 
secondary vertexes are not determined, no angular analysis can be performed. 

In the case of the $D^+\to K^-\pi^+ \mu^+\nu$
decay from FOCUS \cite{massa} the angular distribution was a crucial piece of information.
In this decays there is a 7\% contribution from the $K^-\pi^+$ S-wave. The S-wave
component interferes with the $\overline{K}^{*}(892)^-$,
causing an asymmetry in the helicity angle distribution (the helicity
angle is defined as the angle between the kaon momentum and the line of flight of the $D^+$,
measured in the $K^-\pi^+$ rest frame). The $K^-\pi^+$ line shape from $D^+\to K^-\pi^+ \mu^+\nu$
could be fitted equally well 
with different S-wave models, but each model has a different interference pattern with the P-wave,
distorting the helicity angle distribution  in 
a different way. The distribution of the helicity angle could then be used to discriminate 
between the different S-wave models.

A strong case for the  $\kappa(800)^-$ would be made from the 
$\tau^- \to \overline{K}^0\pi^-\nu_{\tau}$ decay if the angular analysis was performed.
Unfortunately no information on the angular distribution is available.

\subsection{$D^+ \to K^-\pi^+\pi^+$  \hskip .2cm from FOCUS}

Let's now turn to a situation where the $K\pi$ S-wave is largely dominant. The issue here is
the S-wave phase near threshold, which can only be addressed by heavy flavor decays. The
$D^+ \to K^-\pi^+\pi^+$ is a 'golden mode': large branching fraction, easy to be reconstructed,
very low background and with an S-wave contribution amounting to approximately 80\% of the total
decay rate.

\begin{figure}[htb]
\centering
\includegraphics*[width=70mm]{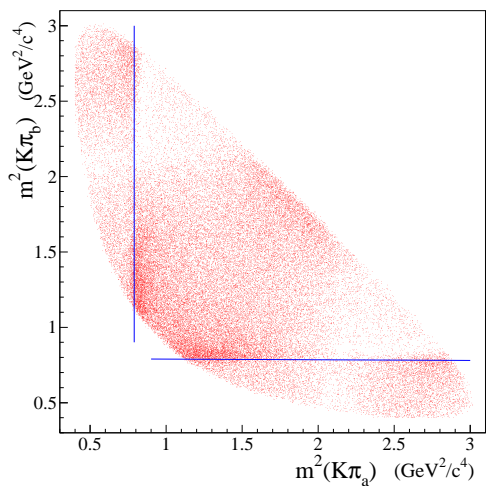}
\caption{The $D^+ \to K^-\pi^+\pi^+$ Dalitz plot from FOCUS. Due to the identical pions, the plot is
symmetric with respect to the diagonal ($m^2(\pi\pi)$) axis.}
\label{kpipidp}
\end{figure}

This decay was studied in great detail by FOCUS \cite{kmat,eu}. A sample with 54K signal events
and 98.5\% purity was analysed with the MIPWA technique. The $D^+ \to K^-\pi^+\pi^+$ Dalitz plot 
is shown in Fig. \ref{kpipidp}. Since there are two identical pions, the Dalitz plot is 
symmetric. The blue lines indicates the $K^{*}(892)$ mass squared. We see clearly the effect
of the angular distribution splitting the $K^{*}(892)$ band into two lobes. A striking feature 
is the displacement of the two lobes with respect to the nominal $K^{*}(892)$ mass. This is readily
explained by the interference between the  $\overline{K}^{*}(892)\pi^+$ mode and the S-wave.
This interference allows one to measure the S-wave phase.

\begin{figure}[htb]
\centering
\includegraphics*[width=70mm]{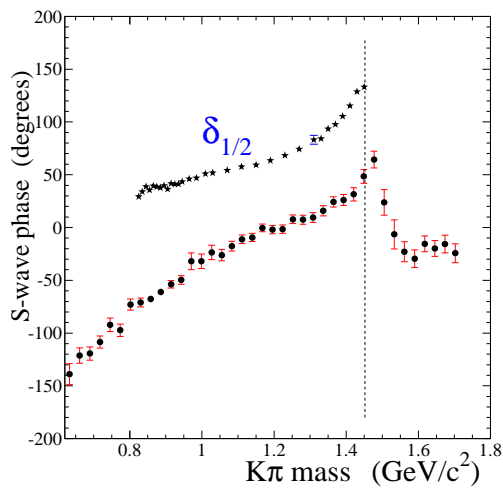}
\caption{The $K^-\pi^+$ MIPWA S-wave (circles with error bars) from FOCUS 
$D^+ \to K^-\pi^+\pi^+$ decay. The LASS $I$=1/2 S-wave phase ($\delta_{1/2}$) is shown as
full stars. The $K\pi$ amplitude is elastic up to 1.45 GeV/$c^2$, indicated by the vertical line.}
\label{e831lass1}
\end{figure}

The S-wave component corresponds to 80.2\% of the decay rate. The remaining part is mostly due to
the P-wave, which is well described by the $\overline{K}^{*}(892)\pi^+$ and the
$\overline{K}^*(1680)\pi^+$ modes (the contribution of the $\overline{K}^*(1410)\pi^+$ mode 
is consistent with zero). A small fraction of $\overline{K}^*_2(1430)\pi^+$ is also present.

\begin{figure}[htb]
\centering
\includegraphics*[width=70mm]{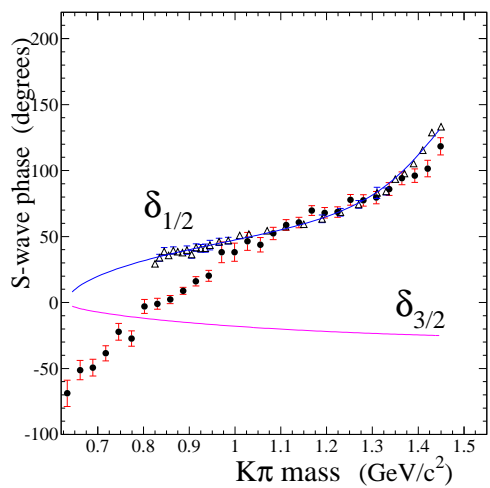}
\caption{The FOCUS $K^-\pi^+$ MIPWA S-wave shifted by 80$^{\circ}$. The lines are the LASS
isospin 1/2 and 3/2 amplitudes. Only the elastic region is shown.}
\label{e831lass2}
\end{figure}

The FOCUS MIPWA S-wave phase is shown in Fig. \ref{e831lass1}. The dotted line indicates the
$K\eta'$ threshold, up to which the $K\pi$ scattering amplitude is elastic. The circles with error
bars are the FOCUS result, whereas the black stars are the LASS $I$=1/2 S-wave phase. The
elastic region is highlighted in Fig. \ref{e831lass2}, which shows  the FOCUS S-wave phase, 
shifted by 80$^{\circ}$, together with the $I$=1/2 and $I$=3/2 S-wave phases from LASS.

All resonances are in the $I$=1/2 component, while the $I$=3/2 amplitude is purely nonresonant.
Since in $D$ decays the nonresonant contribution is usually very small, one would expect
the FOCUS phase to be similar to the LASS $I$=1/2 phase. But we see that this is not the case.
As a matter of fact, no combination of the two LASS isospin components can reproduce the S-wave phase from FOCUS.
The S-wave phases from $D^+ \to K^-\pi^+\pi^+$ and from $K^-\pi^+ \to K^-\pi^+$ are indeed very 
different. An additional energy dependent phase must be added to the LASS phase in order to
match the FOCUS result.

Why are the two phases so different? Where this additional energy dependent phase comes from? 
There are two possible origins:  the decay amplitude and three-body final state interaction. 
It is rather suggestive that the difference between LASS and FOCUS phases increases as one
approaches the $K\pi$ threshold. As the $K\pi$ mass decreases, the momentum of the third 
particle increases, and its interaction with the  $K\pi$  system becomes more intense.
That would explain why the E791 MIPWA S-wave amplitude does not
yield a good fit to the BaBar $D^0 \to K^-K^+\pi^0$ Dalitz plot at low $K\pi$ masses:
at the $K^+\pi^0$ threshold, the third particle (a $K^-$) has smaller momentum, so the
phase introduced by the three-body FSI would be slightly different than that from the
$D^+ \to K^-\pi^+\pi^+$.

The basic problem with the MIPWA S-wave is, therefore, how to interpret the data. The pure
$K\pi$ amplitude is there, covering the entire elastic range from threshold. Unfortunately
it is not directly accessible. We need to learn how to get it.

\section{The $\pi\pi$ amplitude -- the $f_0(1370)$ problem}

The $\pi\pi$ S-wave in the region 1.2-1.5 GeV/$c^2$ is still problematic. There are two
states in this region, namely the $f_0(1370)$ and $f_0(1500)$.
The $f_0(1500)$ is a well established resonance, observed clearly in $pp$ and $p\bar p$ data, and also
in $J/\psi$ decays. Its mass is (1.505$\pm$6) GeV/$c^2$ and its width is
(109$\pm$7) GeV/$c^2$, having also well measured couplings to $\pi\pi$, $4\pi$, $KK$ and 
$\eta\eta$ \cite{pdg}. 

The $f_0(1370)$, by its turn, remains very controversial. Its mass
ranges from 1.2 to 1.5 GeV/$c^2$, while the width lies between 200 and 500 MeV/$c^2$ \cite{pdg}.
The BES Collaboration observed an excess of events in 1.2-1.5 GeV/$c^2$ region in the decay
$J/\psi \to \phi \pi^+ \pi^-$ \cite{besf0}, which was interpreted as a dominant $f_0(1370)$ component
interfering with a small contribution of the $f_0(1500)$. The  $f_0(1370)$ was represented by a
Breit-Wigner and the values obtained for the mass and width were  (1.350$\pm$50) and 
(0.265$\pm$40) GeV/$c^2$, respectively. No evidence of the $f_0(1370)$ was found in 
$J/\psi \to \phi K^+ K^-$ and $J/\psi \to \gamma \pi^+ \pi^-/ \gamma K^+ K^- $. The
ratio of partial widths obtained by BES is consistent with zero:
$\Gamma_{KK}/\Gamma_{\pi\pi}$= (0.08$\pm$0.08).

The region around 1.5 GeV is very interesting: that's where the ground state of the scalar
glueball is expected to be. It is necessary, therefore, to measure not only the $f_0(1370)$
mass and width, but also the couplings to other channels, for this would provide insight to
its nature. One possible scenario includes also the $f_0(1710)$. The three observed states would
be mixtures of two $q \bar q$ and the $0^{++}$ $gg$ states \cite{close}. 

The information given by heavy flavor decays is particularly useful in this respect. The
states that are observed with a large decay fraction in $D$ and $B$ decays are very likely 
to have a dominant $q\bar q$ component.

\subsection{$D_s^+ \to \pi^-\pi^+\pi^+$  \hskip .2cm from FOCUS and E791}

The $D_s^+ \to \pi^-\pi^+\pi^+$ is a 'golden mode' for studies of the $\pi\pi$ system in S-wave.
This is a Cabibbo suppressed mode with no strange quarks in the final state. Resonances that
couple both to $KK$ and to $\pi\pi$, like the $f_0(980)$, are expected to play a dominant role.

\begin{figure}[htb]
\centering
\includegraphics*[width=70mm]{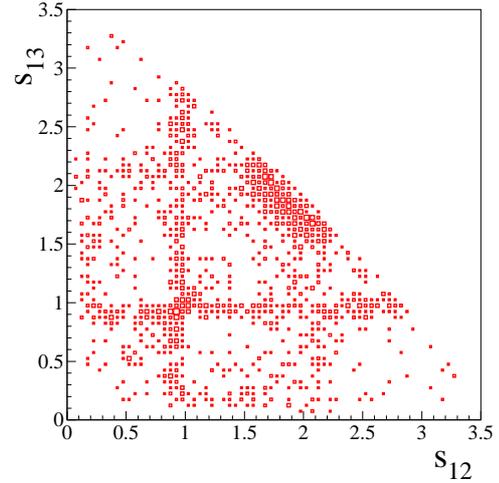}
\caption{The $D_s^+ \to \pi^-\pi^+\pi^+$ Dalitz plot from FOCUS. The narrow bands at 1 GeV/$c^2$
correspond to the $f_0(980)\pi^+$ channel. The concentration of events at 2 GeV/$c^2$ are due
to a scalar state identified with the $f_0(1500)$.}
\label{dsfocus}
\end{figure}

This decay was studied by E791 \cite{ds}, and, more recently, by FOCUS \cite{kmat1}. 
In Fig. \ref{dsfocus} we see the $D_s^+ \to \pi^-\pi^+\pi^+$ Dalitz plot from FOCUS. 
Two features call the
attention immediately: the narrow bands corresponding to the $f_0(980)\pi^+$ mode and the
concentration of events between 1.5-2.2 GeV$^2$/$c^4$. 
This concentration is partially due to a scalar
state with high mass, which will referred to as the $f_0(X)$.

A Dalitz plot analysis was performed both with the K-matrix and with the isobar model 
(but only the result of K-matrix fit were published). In addition to the $f_0(980)$,
the S-wave isobar model included a scalar state, the $f_0(X)$, 
represented by a relativistic Breit-Wigner.
The mass and width of the $f_0(X)$ were determined by the fit.  The values obtained by FOCUS
are (1.476$\pm$6) GeV/$c^2$ for  the mass and (0.119$\pm$18) GeV/$c^2$ for the width. The
data is well described by an S-wave with only two resonances.

\begin{figure}[htb]
\hskip -.3cm \includegraphics*[width=90mm]{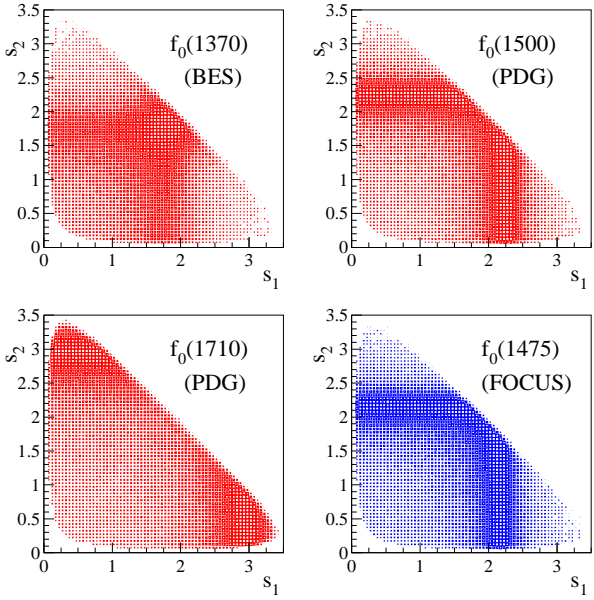}
\caption{Monte Carlo simulation of the Dalitz plot of the $D_s^+ \to f_0(X)\pi^+$ decay, with
different hypothesis for the $f_0(X)$. Clockwise from top left: $f_0(1370)$ (BES), 
$f_0(1500)$ (PDG), $f_0(1710)$ (PDG) and the state found in FOCUS data.}
\label{f0}
\end{figure}

Monte Carlo simulations of the decay $D_s^+ \to f_0(X)\pi^+$ were performed taking
the values obtained by BES and FOCUS for the mass and width of the $f_0(X)$. For comparison, 
simulations were performed assuming for the $f_0(X)$ the PDG values for the $f_0(1500)$ and 
$f_0(1710)$. The simulations are shown in Fig. \ref{f0}. The conclusion is that the scalar
state observed in FOCUS data is much closer to the $f_0(1500)$ than to the state observed
by BES.

\subsection{$D_s^+ \to \pi^-\pi^+\pi^+$  \hskip .2cm from BaBar}

\begin{figure}[htb]
\centering
\includegraphics*[width=85mm]{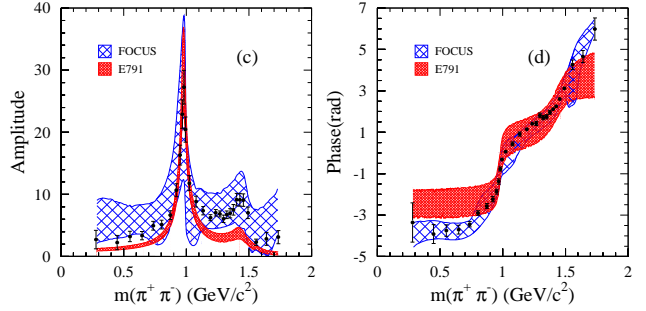}
\caption{Results from the BaBar MIPWA fit of the $D_s^+ \to \pi^-\pi^+\pi^+$ Dalitz plot, showing
the magnitude (left plot) and phase (right plot) of the $\pi^-\pi^+$ S-wave as a function
of the $\pi^-\pi^+$ invariant mass. The two bands superimposed to the BaBar result show
the magnitude and phase of the FOCUS and E791 S-wave form the isobar fit.}
\label{babar3pi}
\end{figure}

Recently BaBar reported study of the $D_s^+ \to \pi^-\pi^+\pi^+$ decay \cite{babar3pi}, from
a high purity sample of about 13000 events. The $\pi^-\pi^+$ S-wave was measured using the 
MIPWA technique. The result is displayed in Fig. \ref{babar3pi}. In the left plot
we see the S-wave magnitude as a function of the $\pi^+\pi^-$mass. There are two peaks, 
a narrow one at the $f_0(980)$ mass and
another at 1.4-1.5 GeV/$c^2$, which is relatively narrow. In the plot on the right we see the 
S-wave phase, also as a function of the $\pi^+\pi^-$mass. There is a rapid variation of the
phase as one crosses the $f_0(980)$ mass, as expected for a typical resonance behaviour. 
The phase continues to grow and
between 1.4-1.5 GeV/$c^2$ another rapid variation can be observed, indicating the presence of
another resonance. The magnitude and phase of the S-wave from the FOCUS analysis is superimposed 
to the BaBar result. The agreement between FOCUS and BaBar is very good.  

Although the conclusion drawn from Fig. \ref{f0} is that the $f_0(X)$ is consistent with the
$f_0(1500)$, the values of the mass and width obtained from the 
$D_s^+ \to \pi^-\pi^+\pi^+$ Dalitz plot fit are not quite the same as the PDG values for the 
$f_0(1500)$. We should keep in mind that the Breit-Wigner which was used is only an approximate 
representation for this state. In FOCUS analysis the $\pi\pi$ mode was assumed to account for 
all the $f_0(1500)$ decay rate. The total decay width should be $\Gamma(s)= \Gamma_{\pi\pi}(s) + 
\Gamma_{4\pi}(s) + \Gamma_{KK}(s) + \Gamma_{\eta\eta}(s)$. Moreover, since the $D_s^+$ mass
is not too high,  the $f_0(1500)$ peak lies out of the Dalitz plot boundary.

\subsection{$B^+ \to K^+\pi^+\pi^-$, $B^0 \to \overline{K}^0\pi^+\pi^-$  
\hskip .2cm from Belle}

Charmless three-body $B$ decays are a very promising tool for light quark spectroscopy.
As in the case of charm decays, charmless $B$ decay have a rich resonant structure. The
phase space of $B$ decays is much larger than that of $D$ decays, so resonances are 
fully contained in the Dalitz plot. However, since the
branching fractions are typically between 10$^{-5}$-10$^{-6}$, the statistics is still limited.
This will no longer be an issue when the LHCb data becomes available. 

\begin{figure}[htb]
\centering
\includegraphics*[width=65mm]{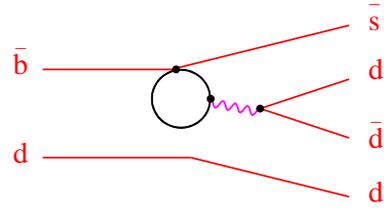}
\vskip -2cm
\caption{The dominant amplitude for the $B^0 \to \overline{K}^0\pi^+\pi^-$ decay. The
diagram for the $B^+ \to K^+\pi^+\pi^-$ is obtained replacing the $d$ quarks by $u$.}
\label{penguin}
\end{figure}

Two decay modes are particularly interesting for the $f_0(1370)$ problem: 
$B^+ \to K^+\pi^+\pi^-$ and $B^0 \to \overline{K}^0\pi^+\pi^-$. The dominant mechanisms in the 
$B^0 \to \overline{K}^0\pi^+\pi^-$ decay are assumed to be the penguin diagram of 
Fig. \ref{penguin} and a tree-level Cabibbo suppressed diagram (external {\em W}-radiation).
The diagrams for $B^+ \to K^+\pi^+\pi^-$ are obtained replacing the $d$ by an  $u$ quark.
One expects, therefore, the same intermediate states, except for the charge of the $K\pi$
resonances, with similar decay fractions.

\vskip -1cm

\begin{figure}[htb]
\centering
\includegraphics*[width=75mm]{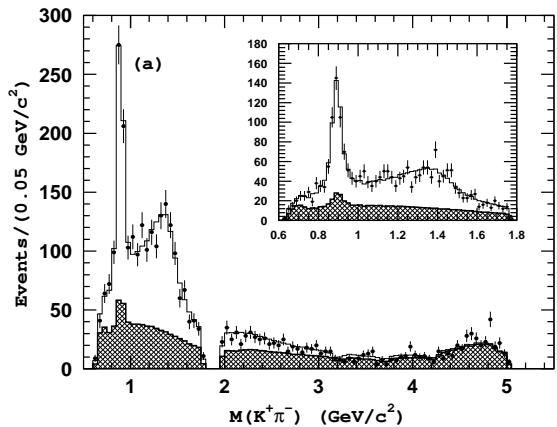}
\caption{The $\pi^+\pi^-$ projection of the $B^+ \to K^+\pi^+\pi^-$ Dalitz plot from Belle.}
\label{bellepipi1}
\end{figure}

\begin{figure}[htb]
\centering
\includegraphics*[width=75mm]{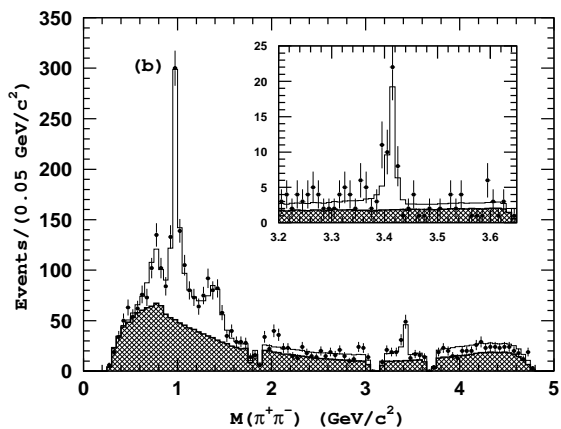}
\caption{The $K^+\pi^-$ projection of the $B^+ \to K^+\pi^+\pi^-$ Dalitz plot from Belle.}
\label{bellepipi2}
\end{figure}

These two modes were studied by Belle \cite{bellekspipi,bellekpipi}. In Figs. \ref{bellepipi1}
and \ref{bellepipi2} we see the projection of the $B^+ \to K^+\pi^+\pi^-$
Dalitz plot onto the $K^+\pi^-$ and $\pi^+\pi^-$ axes, respectively. The  $K^+\pi^-$
projection exhibits two prominent structures, the narrower corresponding to the $K^*(892)\pi^+$
decay and a broader corresponding to the $K_0^*(1430)\pi$.

In Fig.\ref{bellepipi2} three peaks are clearly visible, corresponding to the $\rho(770)$, 
to the $f_0(980)$. The third peak is well described by a model with only one high mass scalar state
at $\sim$1.4-1.5 GeV/$c^2$.

The decay fractions from the Belle Dalitz plot fits are presented  in Table  \ref{b2kpipi}.
The empirical parametrisation $a_1 e^{\delta_1} \ e^{-\alpha s_{K\pi}} + 
a_2 e^{\delta_2} \ e^{-\alpha s_{\pi\pi}}$
was used for the nonresonant amplitude. The nonresonant contribution is dominated by the
$K\pi$ component in both $B^+$ and $B^0$ decays. The very high fraction of the $K_0^*(1430)\pi$ 
fraction is a bit surprising. It is hard to believe that the $K_0^*(1430)\pi$ decay fraction
is five to six times larger than that of the $K^*(892)\pi$. Here we see again the interplay 
between broad structures in the S-wave, resulting in abnormally large 
decay fractions. 

However, when we look to the $\pi\pi$ component, everything seems under control. The peak at
1.4-1.5 GeV/$c^2$ in the $\pi^+\pi^-$ projection is well described by a single scalar resonance,
modeled by a Breit-Wigner amplitude. The Breit-Wigner parameters of this resonance were obtained 
from data:  $M_0 =$(1.449$\pm$0.013) GeV/$c^2$, \ $\Gamma_0 =$(0.126$\pm$0.025) GeV/$c^2$.
These values are in good agreement with the ones from FOCUS.

It seems that in heavy flavor decays only one scalar state is observed in the $\pi\pi$
channel. This state is not consistent with the $f_0(1370)$. It is similar to the
$f_0(1500)$, although the masses and widths obtained from Dalitz plot analysis of $D$ and 
$B$ decays are not quite the same. The difference is not large and could be attributed to the
way the parameters were determined by FOCUS and Belle. In any case, hereafter we will refer to
this scalar resonance as the $f_0(1475)$.

\begin{center}
\begin{table}
\begin{tabular}{ccc}   \hline 
 mode               & $B^+ \to K^+\pi^+\pi^-$ & $B^0 \to \overline{K}^0\pi^+\pi^-$ \\ \hline

$K^*(892)\pi$       & 13.0$\pm$1.0 & 11.8$\pm$1.7  \\   
$K_0^*(1430)\pi$    & 65.5$\pm$4.5 & 64.8$\pm$7.8  \\
$\rho(770)K$        &  7.9$\pm$1.0 & 12.9$\pm$2.0  \\   
$f_0(980)K$         & 17.0$\pm$3.6 & 16.0$\pm$4.2  \\
$f_0(X)K$           & 4.1$\pm$0.9  & 3.7$\pm$2.4   \\
nonresonant   	    & 34.0$\pm$2.7 & 41.9$\pm$5.5  \\  \hline
\end{tabular}
\caption{Decay fractions from the Belle $B \to K\pi\pi$ Dalitz plot fits. The fractions for
both modes are in good agreement, as expected from an isospin symmetry argument.}
\label{b2kpipi}
\end{table}
\end{center}

\subsection{$D^+_s \to K^+K^-\pi^+$  \hskip .2cm from CLEO-c}

One interesting aspect is the large decay fraction of the $f_0(1475)$ observed in the
$D_s^+ \to \pi^-\pi^+\pi^+$. Assuming the main decay mechanism to be the $W$-radiation
(Fig. 1), the large fraction of the $f_0(1475)$ in $D_s^+ \to \pi^-\pi^+\pi^+$ may be
interpreted as an indication of a strong $s\bar s$ component in wave function of this state. 
In this case, a large contribution would also be expected in $D^+_s \to K^+K^-\pi^+$. 

\begin{figure}[htb]
\centering
\includegraphics*[width=70mm]{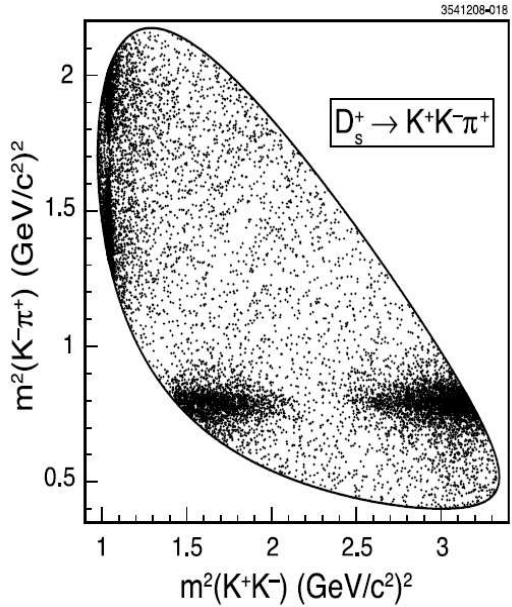}
\caption{The $D_s^+ \to K^+K^-\pi^+$  Dalitz plot from CLEO-c \cite{cleodskkpi}.}
\label{dskkpi}
\end{figure}

The $D^+_s \to K^+K^-\pi^+$ decay was studied by CLEO-c \cite{cleodskkpi}. The CLEO-c sample
has 14K events with very small background. The Dalitz plot is shown in Fig. \ref{dskkpi}.
This is a very tough analysis. The lower part of the $KK$ spectrum is populated by the
$\phi$, the $f_0(980)$ and the $a_0(980)$. There is a strong interference between these amplitudes,
so it is very difficult to separate individual contributions.
We see also the bands corresponding to the
$K^*(892)K^+$ mode. In the region $m_{KK}^2 \sim$2.0-2.2 GeV/$c^2$, however, there is no indication
of a resonance. 

This is very intriguing. Apparently the $f_0(1475)$ does not decay to $KK$. This is
in agreement with the $f_0(1500)$ partial width $\Gamma{KK}/\Gamma{\mathrm{tot}}=0.086\pm0.010$,
and may be considered as an additional evidence for the identification of the two states. But
the mechanism that leads to a large $f_0(1475)$ decay fraction in  
$D_s^+ \to \pi^-\pi^+\pi^+$ remains to be understood.

\section{Conclusions}

\begin{figure}[htb]
\centering
\includegraphics*[width=80mm]{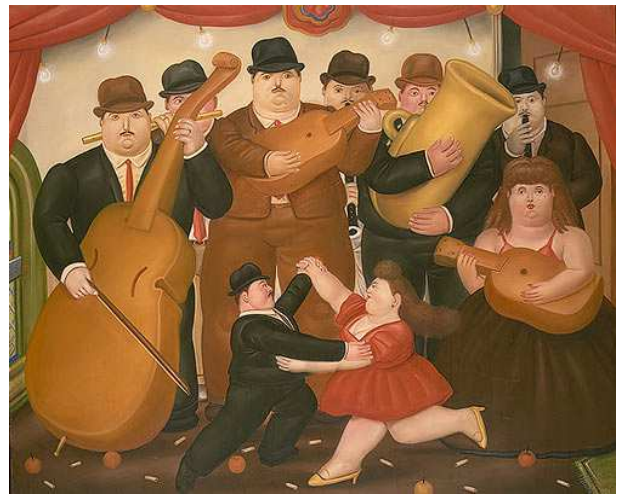}
\caption{Fernando Botero: Dancing in Colombia (1983.251). In Heilbrunn Timeline of Art History. 
New York: The Metropolitan Museum of Art, 2000. 
http://www.metmuseum.org/toah/ho/11/sa/ho-1983.251.htm (October 2006).}
\label{botero}
\end{figure}

The physics of the scalar mesons has still many interesting open problems, which are related
to the strong dynamics at low energy. The main problem in this field is to identify the $q\bar q$ 
mesons of the Constituent Quark Model scalar nonet(s). There are currently more candidates than
slots, although some states remain controversial.  From the experimental point of view, it is not 
trivial to detect broad, structureless overlapping states squeezed in a limited phase space, a situation
which is 
well illustrated by the Colombian painter Fernando Botero, in Fig \ref{botero}. In addition to the
regular $q\bar q$ states, there is a number of other configurations allowed by QCD, like
molecules, hybrids, glueballs, tetraquarks, sharing the same $J^{PC}=0^{++}$ quantum numbers.
These 'exotic' configurations have not been clearly identified yet, but they may be mixed with the
regular  $q\bar q$ mesons.

It is, therefore, unlikely that the understanding of the nature of scalar particles could be
achieved without combining data from different types of reactions. 
Heavy flavor decays have been explored as an
alternative window to some fundamental issues, like the nature of the $\pi\pi$ and $K\pi$ spectrum
near threshold. In this work the $\kappa(800)$ and the $f_0(1370)$ were discussed from the point
of view of hadronic three-body  decays of $D$ and $B$ mesons and also from decays of $\tau$ lepton.

The neutral $\kappa(800)$ is well established. Its pole position was determined, in spite of the lack
of data bellow 825 Mev/$c^2$ in the $K^-\pi^+$ spectrum. This gap can be filled by data from
$D^+ \to K^- \pi^+\pi^+$ using the MIPWA technique, but we need to understand what exactly is being 
measured, how to account for three-body final state interactions, whether or not the
decay dynamics introduces an energy dependent phase.

If the $\kappa(800)$ is a $I$=1/2 state, then its charged partner must exist. The search for the 
$\kappa(800)^{\pm}$ is a hard task, though. The cleanest environment is provided by the semileptonic decays
of $D$ mesons, such as $D \to K\pi\mu\nu$. However, very large samples are required, since the S-wave
is only a small component. Decays of $\tau$ leptons would be an interesting alternative, but the
missing neutrinos is a serious obstacle. Another alternative is to measure the S-wave from decays
like $D^0 \to K^-K^+\pi^0$ (LHCb). The problem here is twofold. In addition to the same difficulties as in the
$D^+ \to K^- \pi^+\pi^+$, in hadron machines it is harder to select a clean
sample of modes with neutral pions. At this point, the nature of the $\kappa$ remains an open question.

The situation concerning the $f_0(1370)$ is also a bit obscure. The existence of this state might not even
be taken for granted. If it had a large $q\bar q$ component, it should have been unambiguously 
observed in charm decays. The scalar state that is present in $D$ and $B$ decays is, instead, similar to the
$f_0(1500)$. The MIPWA analysis of the $D^+_s \to \pi^- \pi^+\pi^+$ show a resonant behaviour between
1.4-1.5 GeV/$c^2$, in agreement with FOCUS and E791 findings. The Breit-Wigner parameters of this
scalar state are not quite compatible with the PDG values for the $f_0(1500)$. One important aspect
that favours the identification of this scalar meson with the $f_0(1500)$ is that it is not seen in 
$D^+_s \to K^- K^+\pi^+$. With more data and a refined analysis technique the pole position of the
$f_0(1475)$ could be determined.

Very soon we will enter the era of 'infinite statistics' in essentially all interesting decay modes. The
expected numbers from LHCb are really impressive. But we must acknowledge that today we are already
limited by systematics. The experimentalists are not ready to explore the full potential of the coming
data. The models currently used to parametrise the signal distributions, like the
Dalitz plot, are inadequate. New analysis tools with a better theoretical foundation are urgently necessary.

\end{document}